\documentstyle[prl,aps,floats,twocolumn,epsf]{revtex}
\newcommand{\beq}{\begin{equation}}
\newcommand{\eeq}{\end{equation}}
\newcommand{\ben}{\begin{eqnarray}}
\newcommand{\een}{\end{eqnarray}}
\newcommand{\bea}{\begin{array}}
\newcommand{\eea}{\end{array}}
\newcommand{\rot}{\omega_{{\rm rot}}}
\newcommand{\corr}{{\rm corr}}
\newcommand{\RPA}{{\rm RPA}}
\newcommand{\Delt}{{\mit \Delta}}

\def \frac#1#2{ { #1 \over #2} }
\def\ket#1{|#1 \rangle}
\def\bra#1{\langle #1 |}

\begin{document}

\draft

\sloppy

\title{\bf On the Response Function Technique for Calculating
          the Random-Phase Approximation Correlation Energy}

\author{ Y.~R.~Shimizu$^1$, P.~Donati$^{2,3}$ and R.~A.~Broglia$^{2,3}$}

\address{
     $^1$Department of Physics, Kyushu University, Fukuoka, 812-8581 Japan.\\
         $^2$Dipartimento di Fisica, Universit\`{a} di Milano and INFN,
          Sez. di Milano, via Celoria 16, 20133 Milano.\\
         $^3$The Niels Bohr Institute, University of Copenhagen,
         Blegdamsvej 17, DK-2100 Copenhagen \O, Denmark.
         }

\date{\today}

\maketitle

\begin{abstract}

We develop a scheme to exactly evaluate the correlation energy
in the random-phase approximation,
based on linear response theory~\cite{PairRMP}.
It is demonstrated that our formula is completely equivalent 
to a contour integral representation recently proposed in Ref.~\cite{Donau}
being numerically more efficient for realistic calculations.
Numerical examples are presented
for pairing correlations in rapidly rotating nuclei.

\end{abstract}


\vspace{1cm}

Mean field theory provides a powerful approximation for describing
many-particle systems.  However, there are many situations
where one is forced to go beyond mean field to include higher-order 
correlations, the random-phase approximation (RPA) being a widely used 
method for this purpose.
One of the basic observables which require to go beyond mean field 
approximation 
is the correlation energy. In this case, the different RPA modes
contribute in a democratic way.  To calculate this quantity one has, 
therefore, to determine
many RPA eigenmodes, especially in the case where symmetries of
the mean-field are spontaneously broken, such as in the case of superfluid 
and deformed rotating nuclei. In Ref.~\cite{PairRMP}
we developed a  method to calculate the correlation
energy  in the RPA, making use of the response function
techniques, and applied it to the study of pairing correlations 
in rapidly rotating nuclei.
The essence of the method consists in expressing
the RPA correlation as an integral
in terms of the RPA response function, function
which can be calculated without explicitly solving the RPA
eigenvalue problem.
These techniques have been recently extended
to deal with the Nambu-Goldstone modes~\cite{PaolaNG},
and to calculate the nucleon effective mass 
in superfluid, deformed, and rotating nuclei~\cite{PaolaSelf}.

Recently, a contour integral representation
for the RPA correlation energy has been proposed in Ref.~\cite{Donau}.
It was claimed that 
this method is more efficient than that developed in Ref.~\cite{PairRMP}.
The assertion was also made that the method of Ref.~\cite{PairRMP}
did not take the contribution of the spurious modes into account.
In this letter we show that both methods are completely equivalent.
Furthermore, using the property of meromorphic functions,
we improve the original method of Ref.~\cite{PairRMP} to obtain the exact 
RPA correlation energy in more efficient way than that proposed 
in~\cite{Donau}.

The Hamiltonian describing the system under discussion is
\beq
      H = H_0 + V,
\label{Hamil}
\eeq
where $H_0$ is the unperturbed one-body (mean-field) Hamiltonian
and $V$ is the residual two-body interaction, which is assumed to be
of the multi-separable form
\beq
     V=-\frac{1}{2} \sum_\rho \chi_\rho Q_\rho Q_\rho,
\label{SepInt}
\eeq
$Q_\rho$ being a one-body hermitian operator while $\chi_\rho$ is the 
strength of the interaction in channel $\rho$.
The associated ground state
energies and state vectors of $H_0$ and $H$ are denoted 
$E_0 $, $\ket{\Phi_0}$ and $E$, $\ket{\Psi}$, respectively.
Turning on the interaction adiabatically,
the correlation energy can be written as~\cite{FW}
\beq
     E_\corr \equiv E - E_0
     = \int_0^1 d\lambda\, \bra{\Psi(\lambda)}V\ket{\Psi(\lambda)}.
\label{Ecorr}
\eeq
In this equation $\ket{\Psi(\lambda)}$ is the ground state of
the $\lambda$-scaled Hamiltonian $H(\lambda) \equiv H_0 + \lambda V$.
Within the RPA approximation,
the correlation energy takes the form (see e.g.~\cite{RS})
\beq
     E_\corr^\RPA = \frac{1}{2}\Bigl[ \sum_n\omega_n
	 - \sum_{\alpha < \beta}(E_\alpha+E_\beta) \Bigr],
\label{EcorrRPA}
\eeq
where $\omega_n$ is the RPA eigenfrequency and
$E_\alpha+E_\beta$ is the unperturbed two-quasiparticle energy
(eigenstates of $H_0$) in the quasiparticle representation.
As shown in~\cite{PairRMP}, the above expression can be rewritten as
\beq
     E_\corr^\RPA = -\frac{1}{2\pi}\,
     \mathop{{\rm lim}}\limits_{\epsilon \rightarrow 0+}\,
       \int_0^\infty d\omega\,
     {\rm Im}\bigl[F(\omega+i\epsilon)\bigr],
\label{EcorrSep1}
\eeq
where
\beq
     F(\omega) \equiv \int_0^1 d\lambda\, {\rm Tr} \Bigl[
       {\cal R}^{(\lambda)}(\omega)\chi \Bigr].
\label{EcorrSep2}
\eeq
The $\lambda$-scaled RPA response function (matrix)
${\cal R}^{(\lambda)}(\omega)$ is defined in term of
the unperturbed response function (matrix),
\beq
     R_{\rho\sigma}(\omega) \equiv \sum_{\alpha < \beta}
   \Bigl[ \frac{q_\rho^*(\alpha\beta)q_\sigma(\alpha\beta)}
      {E_\alpha +E_\beta -\omega}
    + \frac{q_\rho(\alpha\beta)q_\sigma^*(\alpha\beta)}
      {E_\alpha +E_\beta +\omega} \Bigr],
\label{RespSep1}
\eeq
as
\beq
    {\cal R}^{(\lambda)}(\omega)
    = \bigl[1 - R(\omega)\chi\lambda\bigr]^{-1} R(\omega),
\label{RespSep2}
\eeq
where $q_\rho(\alpha\beta)=\bra{\alpha\beta} Q_\rho \ket{0}$
and $\chi = (\delta_{\rho\sigma} \chi_\rho)$.

\begin{figure}[!ht]
\epsfxsize 6cm
\centerline{
\epsfbox{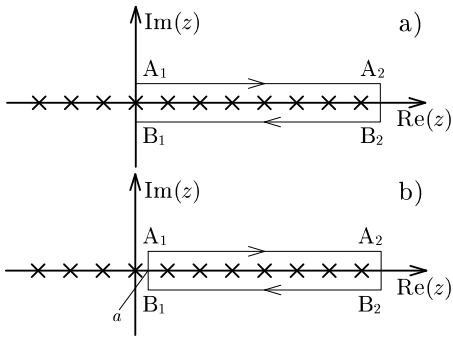}}
\vskip 0.2cm
\caption{
An illustration of the integration contour in the complex plane.
Crosses denote the positions of the RPA and unperturbed roots,
which give singularities of the funtcion $F(z)$,
Eq.~(\protect{\ref{Flogdet}}).
}
\label{contr1}
\end{figure}

A small but finite value of the imaginary part 
$\epsilon \equiv {\rm Im}(\omega)$
has been used to evaluate
the correlation energy in Ref.~\cite{PairRMP} (cf. Eq.~\ref{EcorrSep1}) and 
found that it takes increasingly more computational time to approach the exact
value of $E^{\rm RPA}_{\rm corr}$ ($\epsilon \rightarrow 0$).
Here we show that Eqs.~(\ref{EcorrSep1})$-$(\ref{RespSep2})
are equivalent to the integral representation
proposed in Ref.~\cite{Donau},
and present a more efficient way to evalute the associated correlation energy
than that presented in Ref.~\cite{Donau}.
For this purpose, we make use of the analytic properties of the function
$F(\omega)$, which become evident by carrying out the $\lambda$-integration
in Eq.~(\ref{EcorrSep2}) analytically,
\beq
     F(z)= - \log\bigl(\det \bigl[ 1 - R(z)\chi \bigr]\bigr).
\label{Flogdet}
\eeq
Thus, $F(\omega)$ has logarithmic singularities at the RPA 
eigenenergies
and at the two-quasiparticle unperturbed energies on the real axis.
Eq.~(\ref{EcorrSep1}) can now be expressed as contour integral 
(see Fig.~\ref{contr1})
\beq
     E_\corr^\RPA = - \frac{1}{4\pi i}
	   \oint_{C_{\rm \ref{contr1}a}} dz\, F(z),
\label{EcorrSep3}
\eeq
where $C_{\rm \ref{contr1}a}$ is a (closed) path which passes through the 
origin of the complex plane $z$, and which encloses
all the positive RPA and unperturbed roots.
Even if there is a spurious zero-energy mode
(the symmetry-recovering mode or the Nambu-Goldstone mode)
in the RPA spectrum,
$F(\omega) \sim {\rm const.}\times \log z$ at $z \sim 0$
so that the integral converges at the origin.
In order to show the equivalence of Eq.~(\ref{EcorrSep3}) to
the corresponding formula of Ref.~\cite{Donau}, we consider the 
integration path
dipicted in Fig.~\ref{contr1}b, where $a$ is a small positive
quantity chosen to be smaller than the lowest singular
point of $F(z)$.  Integrating by parts, one obtain
\beq
     \oint_{C_{\rm \ref{contr1}b}} dz\, F(z)
     = \bigl[ z F(z) \bigr]_{C_{\rm \ref{contr1}b}}
    - \oint_{C_{\rm \ref{contr1}b}} dz\, z \frac{d}{dz} F(z).
\label{CCut1}
\eeq
If the zero mode exists, the segment of the real axis
between the origin and the lowest RPA (or two-quasiparitcle) root
is a ``branch-cut'' of the complex logarithmic function $F(z)$,
and so $\bigl[ z F(z) \bigr]_{C_{\rm \ref{contr1}b}} = -2\pi i a$
(minus sign arizing from the fact that the direction of the path
is clockwise).
On the other hand, the singularities of the function $z\frac{d}{dz}F(z)$ are
poles at the same points as those of $F(z)$.
Thus, in the limit $a \rightarrow 0$,
the branch-cut contribution vanishes, and 
\beq
     E_\corr^\RPA = \frac{1}{4\pi i}
	 \oint_{C_{\rm \ref{contr1}b}} dz\, z \frac{d}{dz} F(z),
\label{EcorrSep4}
\eeq
which is nothing else than the integral representation of $E_\corr^\RPA$
of Ref.~\cite{Donau} (cf. Eq.~(6) in it). Note that if there is a zero mode,
$\frac{d}{dz}F(z) \sim 1/z^2$, so the integral in Eq.~(\ref{EcorrSep4})
diverges when the path goes through the origin. Otherwise, the origin
is not a singular point and the path $C_{\rm \ref{contr1}b}$ can be
trivially modified into $C_{\rm \ref{contr1}a}$ in Eq.~(\ref{EcorrSep4}).
If one uses the path $C_{\rm \ref{contr1}b}$
in Eq.~(\ref{EcorrSep3}) instead of $C_{\rm \ref{contr1}a}$,
one has to add the branch-cut contribution $-a/2$
if the zero mode exists.

In spite of the equivalence mentioned above, Eq.~(\ref{EcorrSep3})
is numerically easier to calculate than Eq.~(\ref{EcorrSep4}) 
(or equivalently Eq.~(6) of Ref.~\cite{Donau}), 
since the meromorphic function $F(z)$ has a more regular
asymptotic behavior than $z\frac{d}{dz}F(z)$. In fact,
in the limit $|z| \rightarrow \infty$, $F(z) \rightarrow o(1/z^2)$
while $z\frac{d}{dz}F(z) \rightarrow o(1)$.
Consequently, the contribution to the corresponding integral arising
from the segment A$_2$B$_2$ in Fig.~\ref{contr1} vanishes
when the points A$_2$ and B$_2$ are taken to infinity.
Now the meaning of the approximation used
in Ref.~\cite{PairRMP}, is clear; the calculation of
Eq.~(\ref{EcorrSep1}) with finite $\epsilon \equiv {\rm Im}(\omega)$
is equivalent to the integration along the path shown in Fig.~\ref{contr1}a),
except the contribution from the segment A$_1$B$_1$ which 
vanishes only in the limit of
$\epsilon \equiv {\rm Im}(\omega) \rightarrow 0$.
We have checked this point numerically.

\begin{figure}[!ht]
\epsfysize 4cm
\centerline{
\epsfbox{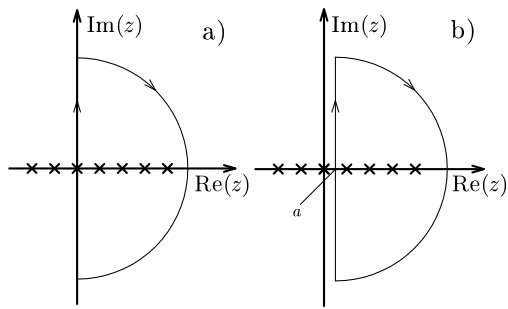}}
\vskip 0.2cm
\caption{
The modified integration contour from Fig.~\protect{\ref{contr1}},
which is more suitable to calculate the RPA correlation energy.
One has to take the limit of infinite radius of the semicircle,
whose contribution then vanishes.
}
\label{contr2}
\end{figure}

For convenience, the integration path in Eq.~(\ref{EcorrSep3})
can be modified to the one shown in Fig.~\ref{contr2}a). In this case
the contribution from the semicircle vanishes
as its radius goes to infinity.
Moreover, the modification of the path
parallel to the real axis into the one
parallel to the imaginary axis is very useful for making
the numerical calculations efficient.
This is because ${\rm Im} F(z)$ is an oscillating
function of ${\rm Re}(z)$ on the path shown in Fig.~\ref{contr1}a),
while ${\rm Re} F(z)$
is a rapidly decreasing function of {\rm Im}(z) on the path shown in
Fig.~\ref{contr2}a). Consequently, one needs in this case
a smaller set of mesh points to carry out
the integration than in the case of the path shown in Fig.~\ref{contr1}a).
Therefore, the expression 
\beq
     E_\corr^\RPA = - \frac{1}{4\pi} \int_{-\infty}^{+\infty} d\omega\,
       {\rm Re} \bigl[F(i\omega)\bigr],
\label{EcorrSep5}
\eeq
of the RPA correlation energy is numerically more convenient to evaluate
than the expression displayed in Eq.~(\ref{EcorrSep3})
or (\ref{EcorrSep4}), equivalent to Eq.~(6) of Ref.~\cite{Donau}.
It is worthwhile noticing that the integral appearing in
Eq.~(\ref{EcorrSep3}) or (\ref{EcorrSep5}), is further simplified by
using the following properties of $F(z)$:
$[F(z)]^*=F(-z^*)$ and $F(-z)=F(z)$, which can be easily demonstrated making 
use of Eqs.~(\ref{EcorrSep1}) and (\ref{Flogdet}).
Consequently the integrand ${\rm Re}F(z)$ is symmetric
with respect to the real axis,
and $\int_{-\infty}^{+\infty} = 2\int_0^{\infty}$ in Eq.~(\ref{EcorrSep5}).
 A similar simplification is possible in Eq.~(\ref{EcorrSep3}).

In Refs.~\cite{PairRMP,RPAvsNP},
the pairing correlations in rapidly rotating nuclei have been studied
by using the general method discussed above.  In these references, in
addition to
the RPA correlation energy, another measure of pairing correlations
was introduced, namely the RPA pairing gap, $\Delt_{\RPA}$~\cite{RPAvsNP}
(called the ``effective'' pairing gap in \cite{PairRMP}).
It is defined as
\beq
  \Delt_{\RPA} \equiv \sqrt{ \Delt^2 + \frac{1}{2}G^2\, S_0({\RPA}) },
\label{DelRPA}
\eeq
with
\beq
  S_0({\RPA}) \equiv \sum_{n \neq {\rm NG}}
    \Bigl[ | \bra{n} P \ket{0} |^2
     + | \bra{n} P^\dagger \ket{0} |^2 \Bigr]_{\RPA },
\label{S0RPA}
\eeq
where $\Delt=G\, \bra{0} P^\dagger \ket{0}_{\rm HB}$
is the standard, static BCS pairing gap 
(the order parameter of mean-field), while
$G$ is the pairing force strength.
The non-energy weighted sum rule $S_0({\RPA})$
describes the contribution of the RPA fluctuations
for the monopole pair transfer operator,
$P^\dagger=\sum_{i>0} c_i^\dagger c_{\tilde i}^\dagger$.
Note that $\sum_{n \neq {\rm NG}}$ means that the divergent contribution
from the spurious mode (pairing rotation) is to be excluded, in keeping with
the fact that its contribution to Eq.~\ref{DelRPA} is included through the 
static pairing gap 
$\Delt$.
In Ref.~\cite{PairRMP}, $S_0({\RPA})$ was calculated making use of the
expression
\beq
  S_0({\RPA}) \approx \frac{1}{\pi} \int_{\omega_{\rm cut}}^\infty d\omega\,
     {\rm Im}\,{\rm Tr}\bigl[{\cal R}(\omega+i\epsilon)\bigr],
\label{S0RPA1}
\eeq
where ${\cal R}(\omega) \equiv {\cal R}^{(\lambda=1)}(\omega)$ is
the RPA response function, whose dimension is 2 corresponding to
$Q_1=(P^\dagger+P)/\sqrt{2}$ and $Q_2=i(P^\dagger-P)/\sqrt{2}$.
A finite value of $\epsilon$ and a low-energy cutoff $\omega_{\rm cut}$
are used to get rid of the NG mode contribution numerically.
This is the same approximation
as that used in calculating the RPA correlation energy~\cite{PairRMP},
and can then be avoided in keeping with the discussion leading to 
Eqs.~(\ref{EcorrSep3}) and (\ref{EcorrSep5}). In this case, the path in 
Fig.~\ref{contr2}b) is to be used in order to avoid the singularity 
associated with an eventual zero mode, as in this case ${\cal R}(z)$ has 
a second order pole at the origin (cf.~\cite{PaolaNG}):
\beq
  S_0({\RPA}) = \frac{1}{2\pi} \int_{-\infty}^{+\infty} d\omega\,
     {\rm Re}\,{\rm Tr}\bigl[{\cal R}(a+i\omega)\bigr].
\label{S0RPA2}
\eeq
Since the function ${\rm Tr}\bigl[ {\cal R}(z) \bigr]$
has poles as singularities, the integral is independent of
the choice of $a$.
Summing up, making use of Eqs.~(\ref{EcorrSep5})
and (\ref{S0RPA2}), both the RPA correlation energy
and the RPA pairing gap can 
be exactly evaluated in a numerically efficient way.

\begin{figure}[!ht]
\epsfysize 7.5cm
\centerline{
\epsfbox{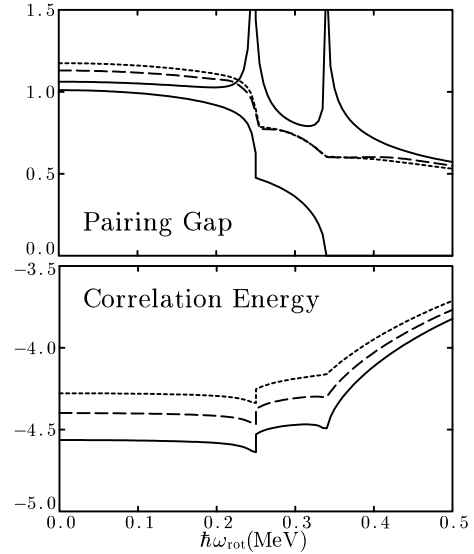}}
\vskip 0.2cm
\caption{
RPA Pairing gap (upper panel) and RPA correlation energy (lower panel)
for neutrons in $^{164}$Er as a function of the rotational frequency.
Both quantities are in MeV.
The solid, dashed and dotted lines denote the results of calculations
with $\epsilon \equiv$ Im($\omega$) = 0 (exact),
100 keV and 200 keV, respectively.
$\hbar \omega_{\rm cut}$ = 400 keV is used
for the approximate (finite $\epsilon$) calculations of the RPA paring gap.
The value of the static (mean-field) pairing gap $\Delta$,
which vanishes at $\hbar \omega_{\rm rot} = 0.34$ MeV,
is also displayed in the upper panel.
}
\label{cdeom}
\end{figure}

In what follows we compare the results of the exact and approximate 
calculations of both $E^{\rm RPA}_{\rm corr}$ and $\Delt_{\rm RPA}$ in 
the case of deformed, superfluid nuclei as a function of the rotational 
frequency (cf. Fig.(3)).
Eqs.~(\ref{EcorrSep5}) and (\ref{S0RPA2}) were used to carry out
the exact calculation, while Eqs.~(\ref{EcorrSep1}) and (\ref{S0RPA1})
with finite values of $\epsilon \equiv {\rm Im}(\omega)$ = 100, 200
keV were used to obtain the approximate results,
as was been done in Refs.~\cite{PairRMP,RPAvsNP}
(${\rm Im}(\omega)$ = 80 keV was used in~\cite{PairRMP,RPAvsNP}).
The nucleus $^{164}$Er has been chosen as a typical rotating nucleus
and constant deformation parameters were used for simplicity.
Only the monopole pairing force has been included with
the smoothed pairing gap method being employed,
which leads to a slightly different model space from
that used in \cite{PairRMP,RPAvsNP}.
Note that the correlation energy given in Eq.~(\ref{EcorrRPA})
as well as the sum rule value of the pairing gap given in Eq.~(\ref{S0RPA})
include the exchange (Fock) contribution.  In Fig.~\ref{cdeom},
this contribution is excluded for the RPA pairing gap in accordance with
Ref.~\cite{RPAvsNP}, while it is included for the RPA correlation
energy as in Ref.~\cite{PairRMP}.
The cusp behaviours are clearly visible in the correlation energy,
which are caused by the sudden change of the mean-field (static pairing gap);
they are associated with the pairing phase-transition at $\hbar\rot=0.34$ MeV,
and
with the crossing of the $g$- and $s$-bands at $\hbar\rot=0.25$ MeV.
They are more evident in the RPA pairing gap,
where the ``exact'' ($\epsilon=0$) calculation diverges at these
transition points.
These divergences are due to a contribution of the lowest solution,
which approaches to zero-energy at the transition points
and brings about similar effects as those caused by
the spurious mode, e.g. infinite strength.
This is the well-known drawback of the RPA, whose small amplitude
approximation breaks down near the transition points.
The approximate results ($\epsilon\ne 0$) are very similar to the exact one
($\epsilon= 0$)
and provide accurate estimates
of the rotational frequency dependence of both the correlation energy
and the pairing gap, except
at the transition points. In particular in the case of the pairing gap,
the singular behaviour 
at these points are smooth out because of
the low-energy cutoff $\hbar\omega_{\rm cut}$.

\begin{figure}[!ht]
\epsfysize 7.5cm
\centerline{
\epsfbox{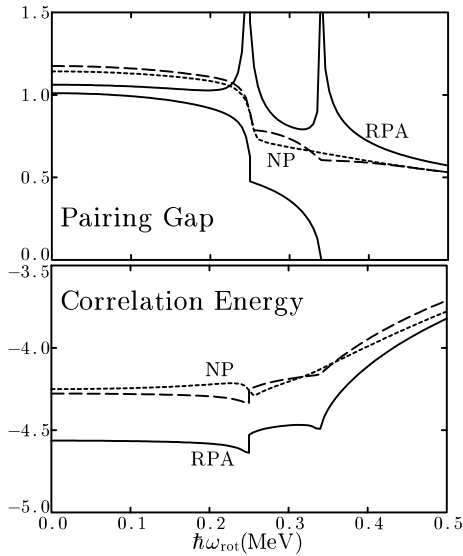}}
\vskip 0.2cm
\caption{
Comparison of the RPA and the number projection (NP) calculations
of the pairing gap (upper panel) and correlation energy (lower panel)
for neutrons in $^{164}$Er as a function of the rotational frequency.
The solid (dotted) lines denote the results of the RPA (NP) calculations.
The results of approximate $\epsilon=200$ keV calculation
in Fig.~\protect{\ref{cdeom}} are also included as dashed lines.
}
\label{cdenp}
\end{figure}

There is an another method which allows
to go beyond mean-field approximation, namely
number-projection (NP) (see e.g.~\cite{RS}).
The results of both methods were compared in Ref.~\cite{RPAvsNP},
where the RPA calculation was carried out approximately as in~\cite{PairRMP}.
Here we compare the NP results  with the exact RPA results
in Fig.~\ref{cdenp}. The NP correlation energy
is defined as the energy difference between the NP and mean-field
(Hartree-Bogoliubov),
$E^{({\rm NP})}_{\rm corr} \equiv E_{\rm NP} - E_{\rm HB}$
(the exchange energy is included in $E_{\rm NP}$).
Although RPA leads to larger values of the correlations, especially in the 
superfluid phase, the rotational frequency dependences are quite similar
as already found in~\cite{RPAvsNP}.
The advantage of the NP method over the RPA is to lead to smooth functions
for both the correlation energy and the pairing gap
at the pairing phase-transition point.  It is, however, noticed that
the cusp behaviour remains in the correlation energy
at the $g$-$s$ crossing point.

In conclusion, a method to exactly deal with RPA correlations, based on that
proposed in Ref.~\cite{PairRMP}, has been developed. It is equivalent to
a recently proposed integral representation in~\cite{Donau},
and allows one to calculate the exact RPA correlation energy 
and in a numerically efficient way, properly dealing with the contribution of 
the Nambu-Goldstone modes. Making use of this method the
pairing correlation in rapidly rotating nuclei
can be studied in detail, providing the basis for an eventual analysis of the 
pairing phase-transition in strongly rotating nuclei.

   This work is support in part by the Grant-in-Aid for
Scientific Research from the Japan Ministry of Education,
Science and Culture (No. 10640275).


\begin{thebibliography}{99}

\bibitem{PairRMP}
Y.~R.~Shimizu, J.~D.~Garrett, R.~A.~Broglia, M.~Gallardo and E.~Vigezzi, 
Rev. Mod. Phys. {\bf 61}, 131 (1989).
\bibitem{PaolaNG}
P.~Donati, T.~D\o ssing, Y.~R.~Shimizu, P.~F.~Bortignon and R.~A.~Broglia,
Nucl. Phys. {\bf A653}, 27 (1999).
\bibitem{PaolaSelf}
P.~Donati, T.~D\o ssing, Y.~R.~Shimizu, P.~F.~Bortignon and R.~A.~Broglia,
Nucl. Phys. {\bf A653}, 225 (1999).
\bibitem{Donau}
F.~D\"onau, D.~Almehed and R.~G.~Nazmitdinov,
Phys. Rev. Lett. {\bf 83}, 280 (1999).
\bibitem{FW}
A.~L.~Fetter and J.~D.~Walecka,
{\it Quantum Theory of Many-Particle Systems},
(McGraw-Hill, New York, 1971).
\bibitem{RS}
P.~Ring and P.~Schuck,
{\it The Nuclear Many-Body Problem},
(Springer-Verlag, New York, 1980).
\bibitem{RPAvsNP}
Y.~R.~Shimizu and R.~A.~Broglia,
Nucl. Phys. {\bf A515}, 28 (1990);

\end{thebibliography}
\end{document}